\documentclass[letterpaper, 10 pt, conference]{ieeeconf}  

\IEEEoverridecommandlockouts                              
\usepackage{graphicx} 
\usepackage{subcaption}
\usepackage{enumerate}

\usepackage{fancyhdr}

\fancypagestyle{first}{
   \fancyhead[R]{}
   \fancyhead[L]{}
   \fancyfoot[L]{{\scriptsize \textbf{ \textcopyright \the\year{} IEEE. Personal use of this material is permitted. Permission from IEEE must be obtained for all other uses, in any current or future media, including reprinting/republishing this material for advertising or promotional purposes, creating new collective works, for resale or redistribution to servers or lists, or reuse of any copyrighted component of this work in other works.} }}
   \fancyfoot[R]{}
   \fancyfoot[C]{}
   
   }

\title{\LARGE \bf 
Bridging Autoencoders and Dynamic Mode Decomposition for Reduced-order Modeling and Control of PDEs
}

\author{Priyabrata Saha and Saibal Mukhopadhyay
\thanks{The authors are with School of Electrical and Computer Engineering,
        Georgia Institute of Technology, Atlanta, USA. \hspace{5mm}
        {\tt Emails: priyabratasaha@gatech.edu, saibal.mukhopadhyay@ece.gatech.edu}}%
}

\usepackage{bm,amsmath,amssymb}

\def\1{\bm{1}}

\def\vu{{\bm{u}}}

\def\vx{{\bm{x}}}

\def\vz{{\bm{z}}}

\def\mA{{\bm{A}}}
\def\mB{{\bm{B}}}
\def\mC{{\bm{C}}}
\def\mD{{\bm{D}}}
\def\mE{{\bm{E}}}

\def\mG{{\bm{G}}}

\def\mI{{\bm{I}}}

\def\mK{{\bm{K}}}

\def\mU{{\bm{U}}}
\def\mV{{\bm{V}}}

\def\mY{{\bm{Y}}}

\DeclareMathAlphabet{\mathsfit}{\encodingdefault}{\sfdefault}{m}{sl}
\SetMathAlphabet{\mathsfit}{bold}{\encodingdefault}{\sfdefault}{bx}{n}

\def\sI{{\mathbb{I}}}

\def\sR{{\mathbb{R}}}

\def\sU{{\mathbb{U}}}

\def\sW{{\mathbb{W}}}
\def\sX{{\mathbb{X}}}

\newcommand{\R}{\mathbb{R}}

\newcommand{\greek}[1]{\bm{\mathit{#1}}}
\newcommand{\eqdef}{\overset{\Delta}{=}}
\newcommand{\vecm}{\text{vec}}
\newcommand{\rank}{\text{rank}}

\newtheorem{theorem}{Theorem}
\newtheorem{corollary}{Corollary}[theorem]
\newtheorem{lemma}[theorem]{Lemma}

\newtheorem{remark}{Remark}

\begin{document}

\maketitle

\thispagestyle{first}

\begin{abstract}

Modeling and controlling complex spatiotemporal dynamical systems driven by partial differential equations (PDEs) often necessitate dimensionality reduction techniques to construct lower-order models for computational efficiency. This paper explores a deep autoencoding learning method for reduced-order modeling and control of dynamical systems governed by spatiotemporal PDEs. We first analytically show that an optimization objective for learning a linear autoencoding reduced-order model can be formulated to yield a solution closely resembling the result obtained through the \textit{dynamic mode decomposition with control} algorithm. We then extend this linear autoencoding architecture to a deep autoencoding framework, enabling the development of a nonlinear reduced-order model. Furthermore, we leverage the learned reduced-order model to design controllers using stability-constrained deep neural networks. Numerical experiments are presented to validate the efficacy of our approach in both modeling and control using the example of a reaction-diffusion system. 

\end{abstract}

\section{Introduction}
\label{sec:introduction}
Performing high-fidelity simulations of physical systems governed by partial differential equations (PDEs) incurs substantial computational costs, rendering subsequent tasks, such as control, extremely challenging if not infeasible.
To address this, reduced-order models (ROMs) are often developed using dimensionality reduction techniques, enabling efficient simulation and control.
For controlled dynamical systems, the reduced-order modeling approaches either combine analytical techniques with empirical approximation \cite{willcox2002balanced} or are purely data-driven \cite{juang1985eigensystem, juang1993identification, proctor2016dynamic}. Among these, the dynamic mode decomposition (DMD) based methods have become widely popular in recent years due to a strong connection between DMD and Koopman operator theory \cite{rowley2009spectral}. 
Another recent research trend involves utilizing deep neural networks (DNNs), particularly autoencoders, for modeling and control of high-dimensional dynamical systems. 
Most research in this area focuses only on the modeling and prediction of such complex dynamics \cite{lusch2018deep, eivazi2020deep, vlachas2022multiscale}.

A second line of research, though relatively less prevalent than modeling and prediction, leverages deep learning for controlling PDE-driven systems. 
Deep reinforcement learning (RL) is one such approach utilized to learn control policies for these systems \cite{vignon2023recent, wang2024control}.  
However, model-free RL methods require running numerical solvers in every iteration to provide feedback to the agents, which is computationally expensive. The same concern arises for the methods involving differentiable simulators as in \cite{holl2020learning, takahashi2021differentiable}. 
The alternative to model-free methods for control design takes the traditional approach: develop a model first and then use that to design controllers \cite{morton2018deep, bounou2021online, chen2021deep}. 
These model-based methods constrain the latent dynamic models to be linear and work well within a short time-window. However, linear combination of a finite number of dynamic modes may not accurately represent the long-term nonlinear characteristics of complex dynamics \cite{khodkar2019koopman} and adding nonlinear forcing terms yields better prediction accuracy \cite{eivazi2020deep}.
It is also necessary to update the linear ROMs with online observations during operation for better prediction accuracy. 
Accordingly, model-based approaches typically leverage the model-predictive control (MPC) scheme to optimize the control policy online using the updated ROMs.
Running online optimization at each step may not be computationally feasible in many cases.

In this paper, we explore the potential of nonlinear ROMs to achieve more accurate long-term predictions, enabling the use of offline control learning methods. Specifically, we develop autoencoder-based ROMs for PDE-driven controlled dynamical systems and leverage these ROMs to learn control policies for the original systems. Our main contributions are:
\begin{itemize}
    \item Drawing inspiration from the \textit{dynamic mode decomposition with control} (DMDc) algorithm \cite{proctor2016dynamic}, we propose an autoencoding framework that effectively captures the dynamic characteristics of the underlying systems, which is crucial for the controller's performance on the full-order system. 
    \item We formulate an objective function for data-driven model learning in a linear autoencoding configuration and show analytically that this function encourages a linear ROM resembling the model obtained via DMDc.
    \item We configure the linear autoencoding architecture such that its components can be substituted with DNNs and optimized via gradient descent to obtain a nonlinear ROM. We show that DNN-based nonlinear ROMs yield more accurate long-term predictions and facilitates offline control learning for the underlying system.
    \item We leverage the learned nonlinear ROM to design controllers using stability-constrained DNNs, ensuring the stability of the control policies.
\end{itemize}
Section \ref{sec:prelims} provides the problem statement and background on DMDc. Section \ref{sec:method} details the linear autoencoding framework, its connection to DMDc, extends it to nonlinear autoencoders, and outlines the control learning algorithm. Section \ref{sec:results} presents numerical experiments to validate the proposed methods, and Section V concludes the paper with a summary of findings and potential future research directions.

\section{Problem and Preliminaries}
\label{sec:prelims}

\subsection{Problem statement}
Consider a time-invariant controlled dynamical system driven by a PDE
\begin{equation}
    \frac{\partial \mathcal{X}}{\partial t} = \mathcal{M}\Big(\mathcal{X}, \frac{\partial \mathcal{X}}{\partial \greek{\zeta}}, \frac{\partial^2 \mathcal{X}}{\partial \greek{\zeta}^2}, \cdots, \vu\Big), 
    \label{eqn:pde_form}
\end{equation}
where $\mathcal{X}(\greek{\zeta}, t) \in \sR$ is the system state at location $\greek{\zeta}$ and time $t$, and $\vu(t) \in \sU \subset \R^{d_\vu}$ is the actuation (or control input) at time $t$. $\mathcal{M}$ is a linear or nonlinear function of the system state, its spatial derivatives of different orders, and the actuation.
Space discretization of the state 
leads to a system of ordinary differential equations (ODEs) that can be written as  
\begin{equation}
    \frac{d \vx}{d t} = f(\vx , \vu). 
    \label{eqn:fom}
\end{equation}
Here $\vx(t) \in \sX \subset \R^{d_\vx}, d_\vx >> 1 $ 
is the space-discretized state vector
at time $t$. 
Our objective is to learn a reduced-order model for this high-dimensional ($d_\vx >> 1$) system of (\ref{eqn:fom}) and use that ROM to learn a feedback controller $\vu = \pi(\vx)$ that stabilizes the system at a desired state. We consider a data-driven learning scenario and 
assume that we have observations from the system consisting of time series data $\vx(t_i), i = 0, 1, \cdots, n$ subjected to random values of actuations $\vu(t_i), i = 0, 1, \cdots, (n-1)$. Note, we use $v$ (in place of $v(t)$ for brevity) as notation for any continuous-time variable (e.g., system state, control input), whereas $v(t_i)$ is used to denote their discrete sample at time instance $t_i$. 
We further assume that the system we are aiming to stabilize at an equilibrium point is \textit{locally stabilizable}, i.e., there exists a control policy such that the desired state is \textit{asymptotically stable} for the closed-loop system. 

\subsection{Dynamic mode decomposition with control}

\label{subsec:MOR-DMDc}
DMD \cite{schmid2010dynamic} is a data-driven method that reconstructs the underlying dynamics using only a time series of snapshots from the system. 
DMDc \cite{proctor2016dynamic} is an extension of DMD for dynamical systems with control. DMDc seeks best-fit linear operators $\mA$ and $\mB$ between successive observed states and the actuations:
\begin{equation}
    \hat{\vx}(t_{i+1}) = \mA \vx(t_i) + \mB \vu(t_i), \quad i = 0, 1, \cdots , n-1, 
    \label{eqn:dmd-fom}
\end{equation}
where $\hat{\vx}(t)$ denotes an approximation of $\vx(t)$, $\mA \in \R^{d_\vx \times d_\vx}$, and $\mB \in \R^{d_\vx \times d_\vu}$. Direct 
analysis of (\ref{eqn:dmd-fom}) could be computationally prohibitive for $d_\vx >> 1$. DMDc leverages dimensionality reduction to compute a ROM
\begin{subequations}
    \begin{align}
        \vx_{\text{R,DMDc}}(t_i) &= \mE_{\text{DMDc}} \vx(t_i), \\
        \vx_{\text{R,DMDc}}(t_{i+1}) &= \mA_{\text{R,DMDc}} \vx_{\text{R,DMDc}}(t_i) + \mB_{\text{R,DMDc}} \vu(t_i), \\ i &= 0, 1, \cdots , n-1, \nonumber
    \end{align}
    \label{eqn:dmd-rom}
\end{subequations}
which retains the dominant dynamic modes of (\ref{eqn:dmd-fom}). Here, $\vx_{\text{R,DMDc}}(t_i) \in \R^{r_\vx}$ is the reduced state, where $r_\vx << d_\vx$, and  $\mE_{\text{DMDc}} \in \R^{r_\vx \times d_\vx}$, $\mA_{\text{R,DMDc}} \in \R^{r_\vx \times r_\vx}, \mB_{\text{R,DMDc}} \in \R^{r_\vx \times d_\vu}$. The full state is reconstructed from the reduced state using the transformation $\hat{\vx}(t_i) = \mD_{\text{DMDc}} \vx_{\text{R,DMDc}}(t_i)$, where $\mD_{\text{DMDc}} \in \R^{d_\vx \times r_\vx}$.
DMDc computes truncated singular value decomposition (SVD) of the data matrices $\mY = [\vx(t_1), \vx(t_2), \cdots, \vx(t_n)] \in \R^{d_\vx \times n}$ and $\greek{\Omega} = [\bm{\omega}(t_0), \bm{\omega}(t_1), \cdots, \bm{\omega}(t_{n-1})] \in \R^{(d_\vx + d_\vu) \times n}$, $\bm{\omega}(t_i) = [\vx(t_i)^\top, \vu(t_i)^\top]^\top \in \R^{d_\vx + d_\vu}$ as follows:
\begin{equation}
    \mY = \widehat{\mU}_\mY \greek{\widehat{\Sigma}}_\mY \widehat{\mV}_\mY^\top, \quad \greek{\Omega} = \widehat{\mU}_{\greek{\Omega}} \greek{\widehat{\Sigma}}_{\greek{\Omega}} \widehat{\mV}_{\greek{\Omega}}^\top,
    \label{eqn:svd}
\end{equation}
where $\widehat{\mU}_\mY \in \R^{d_\vx \times r_\vx}, \greek{\widehat{\Sigma}}_\mY \in \R^{r_\vx \times r_\vx}, \widehat{\mV}_\mY \in \R^{n \times r_\vx}, \widehat{\mU}_{\greek{\Omega}} \in \R^{(d_\vx+d_\vu) \times r_{\vx\vu}}, \greek{\widehat{\Sigma}}_{\greek{\Omega}} \in \R^{r_{\vx\vu} \times r_{\vx\vu}}$, and $\widehat{\mV}_{\greek{\Omega}} \in \R^{n \times r_{\vx\vu}}$. Here, $r_\vx < \min(d_\vx, n)$ and $r_\vx < r_{\vx\vu} < \min(d_\vx+d_\vu, n)$ denote the truncation dimensions of SVDs. Utilizing the SVDs of (\ref{eqn:svd}), the parameters of the ROM (\ref{eqn:dmd-rom}) is obtained as 
\begin{subequations}
\begin{align}
    \mE_{\text{DMDc}} &= \widehat{\mU}_\mY^\top, \quad  \mD_{\text{DMDc}} = \widehat{\mU}_\mY, \\  \mA_{\text{R,DMDc}} &= \widehat{\mU}_\mY^\top \mY \widehat{\mV}_{\greek{\Omega}} \greek{\widehat{\Sigma}}_{\greek{\Omega}}^{-1} \widehat{\mU}_{\greek{\Omega}, 1}^{\top} \widehat{\mU}_\mY, \\  \mB_{\text{R,DMDc}} &= \widehat{\mU}_\mY^\top \mY \widehat{\mV}_{\greek{\Omega}} \greek{\widehat{\Sigma}}_{\greek{\Omega}}^{-1} \widehat{\mU}_{\greek{\Omega},2}^{\top}, 
\end{align}
\label{eqn:dmdc-params}
\end{subequations}
where $\widehat{\mU}_{\greek{\Omega},1} \in \R^{d_\vx \times r_{\vx\vu}}, \widehat{\mU}_{\greek{\Omega},2} \in \R^{d_\vu \times r_{\vx\vu}}$, and $\widehat{\mU}_{\greek{\Omega}}^{\top} = [\widehat{\mU}_{\greek{\Omega},1}^{\top} \ \  \widehat{\mU}_{\greek{\Omega},2}^{\top}]$.
\section{Method}
\label{sec:method}   

\subsection{Learning a reduced order model}
\label{subsec:rom}
To develop a nonlinear ROM utilizing DNNs that effectively capture the underlying dynamics, we first investigate if we can obtain a linear ROM similar to DMDc, in a gradient descent arrangement. Specifically, we analyze optimization objectives that encourage a DMDc-like solution for a reduced-order modeling problem using linear networks (single layer network without nonlinear activation). Consider the following reduced-order modeling problem
\begin{subequations}
\begin{align}
    \vx_{\text{R}}(t_i) &= \mE_\vx \vx(t_i), \\ 
    \vx_{\text{R}}(t_{i+1}) &= \mA_{\text{R}} \vx_{\text{R}}(t_i) + \mB_{\text{R}} \vu(t_i), \\ \hat{\vx}(t_i) &= \mD_\vx \vx_{\text{R}}(t_i), \  i = 0, 1, \cdots , n-1, 
\end{align}
\label{eqn:dmdc-like}
\end{subequations}
where the linear operators $\mE_\vx \in \R^{r_\vx \times d_\vx}$ and $\mD_\vx \in \R^{d_\vx \times r_\vx}$ projects and reconstructs back, respectively, the high-dimensional system state to and from a low-dimensional feature $\vx_{\text{R}} \in \R^{r_\vx}$. The linear operators $\mA_{\text{R}} \in \R^{r_\vx \times r_\vx}$ and $\mB_{\text{R}} \in \R^{r_\vx \times d_\vu}$ describe the relations between successive reduced states and actuations. We refer to this reduced-order model with linear networks as \textit{linear autoencoding ROM} or LAROM. In the following, we first analyze the solution of the optimization objective of LAROM for a fixed \textit{encoder} $\mE_\vx$. Then we establish a connection between the solution of LAROM and the solution of DMDc, and further discuss the choice of the encoder to promote similarity between the two. Finally, we extend the linear model to a DNN-based model, which we refer to as DeepROM.

\subsubsection{Analysis of the linear reduced-order model for a fixed encoder}
The DMDc algorithm essentially solves for $\widetilde{\mG} \in \R^{r_\vx \times (d_\vx+d_\vu)}$ to minimize $\frac{1}{n} \sum_{i=0}^{n-1} \big\|\mE_\vx \vx(t_{i+1}) - \widetilde{\mG} \bm{\omega}(t_i) \big\|^2$ for a fixed projection matrix $\mE_\vx = \mE_{\text{DMDc}} = \widehat{\mU}_\mY^\top$. Here, $\bm{\omega}(t_i)$ is the concatenated vector of state and actuation as defined in section \ref{subsec:MOR-DMDc}. The optimal solution $\widetilde{\mG}_{\text{opt}}$ is then partitioned as $[\widetilde{\mA} \ \  \widetilde{\mB}]$ such that $\widetilde{\mA} \in \R^{r_\vx \times d_\vx}, \widetilde{\mB} \in \R^{r_\vx \times d_\vu}$.  Finally, $\widetilde{\mA}$ is post-multiplied with the reconstruction operator $\mD_{\text{DMDc}} = \widehat{\mU}_\mY$ to get the ROM components $\mA_{\text{R,DMDc}}$ and $\mB_{\text{R,DMDc}}$.
Note, that the final step of this process (multiplication of the operators) is feasible
only for the linear case, not in the case when the projection and reconstruction operators are nonlinear (e.g. DNNs). Therefore, we use an alternative formulation with the following results to design a loss function that encourages a DMDc-like solution for (\ref{eqn:dmdc-like}) and also offers dimensionality reduction when nonlinear components are used. 

\begin{theorem}
Consider the following objective function 
\begin{equation}
    L_{\normalfont\text{pred}} (\mE_\vx, \mG) = \frac{1}{n} \sum_{i=0}^{n-1} \big\|\mE_\vx \vx(t_{i+1}) - \mG \mE_{\vx \vu} \bm{\omega}(t_i) \big\|^2,
    \label{eqn:lin_loss_pred}
\end{equation} 
where $\mG = [\mA_{\normalfont\text{R}} \ \  \mB_{\normalfont\text{R}}] \in \R^{r_\vx \times (r_\vx+d_\vu)}, \mE_{\vx \vu} = \begin{bmatrix}
\mE_\vx & \mathbf{0}\\
\mathbf{0} & \mI_{d_\vu}
\end{bmatrix} \in \R^{(r_\vx + d_\vu) \times (d_\vx + d_\vu)}$, $\mI_{d_\vu}$ being the identity matrix of order $d_\vu$.
For any fixed matrix $\mE_\vx$, the objective function $L_{\normalfont\text{pred}}$ is convex in the coefficients of $\mG$ and attains its minimum for any $\mG$ satisfying 
\begin{equation}
    \mG \mE_{\vx \vu} \greek{\Omega} \greek{\Omega}^\top \mE_{\vx \vu}^\top = \mE_\vx \mY \greek{\Omega}^\top \mE_{\vx \vu}^\top, 
    \label{eqn:lin_loss_pred_sol_1}
\end{equation}
where $\mY$ and $\greek{\Omega}$ are the data matrices as defined in section \ref{subsec:MOR-DMDc}.
If $\mE_\vx$ has full rank $r_\vx$, and $\greek{\Omega} \greek{\Omega}^\top$ is non-singular, then $L_{\normalfont\text{pred}}$ is strictly convex and has a unique minimum for
\begin{equation}
    \mG = [\mA_{\normalfont\text{R}} \ \  \mB_{\normalfont\text{R}}] = \mE_\vx \mY \greek{\Omega}^\top \mE_{\vx \vu}^\top (\mE_{\vx \vu} \greek{\Omega} \greek{\Omega}^\top \mE_{\vx \vu}^\top)^{-1}.
    \label{eqn:lin_loss_pred_sol_2}
\end{equation}
\label{theorem:lin_opt}
\end{theorem}

\textit{Proof}. See appendix \ref{apdx:proof_lin_opt}.

\begin{remark} 
For a unique solution, we assume that $\mE_\vx$ has full rank. The other scenario, i.e., $\mE_\vx$ is rank-deficient suggests poor utilization of the hidden units of the model. In that case, the number of hidden units (which represents the dimension of the reduced state) should be decreased. The assumption that the covariance matrix $\greek{\Omega} \greek{\Omega}^\top$ is invertible can be ensured when $n \geq d_\vx + d_\vu$, by removing any linearly dependent features in system state and actuation. When $n < d_\vx + d_\vu$, the covariance matrix  $\greek{\Omega} \greek{\Omega}^\top$ is not invertible. However, similar results can be obtained by adding $\ell_2$ regularization (for the coefficients/entries of $\mG$) to the objective function. 
\end{remark}

\subsubsection{The connection between the solutions of the linear autoencoding model and DMDc}
\label{subsub:connection}
The connection between the ROM obtained by minimizing $L_{\text{pred}} $ (for a fixed $\mE_\vx$), i.e., (\ref{eqn:lin_loss_pred_sol_2}) and the DMDc ROM of (\ref{eqn:dmdc-params}) is not readily apparent. To interpret the connection, we formulate an alternative representation of (\ref{eqn:lin_loss_pred_sol_2}) utilizing the SVD and the Moore-Penrose inverse of matrices. This alternative representation leads to the following result. 

\begin{corollary}
Consider the (full) SVD of the data matrix $\greek{\Omega}$ given by $\greek{\Omega} = \mU_{\greek{\Omega}} \greek{\Sigma}_{\greek{\Omega}} \mV_{\greek{\Omega}}^\top$, where $\mU_{\greek{\Omega}} \in \R^{(d_\vx+d_\vu) \times (d_\vx+d_\vu)}, \greek{\Sigma}_{\greek{\Omega}} \in \R^{(d_\vx+d_\vu) \times n}$, and $\mV_{\greek{\Omega}} \in \R^{n \times n}$. If $\mE_\vx = \widehat{\mU}_\mY^\top$ and $\greek{\Omega} \greek{\Omega}^\top$ is non-singular, then the solution for $\mG = [\mA_{\normalfont\text{R}} \ \  \mB_{\normalfont\text{R}}]$ corresponding to the unique minimum of $L_{\text{pred}} $ can be expressed as
\begin{equation}
    \mA_{\text{R}} = \widehat{\mU}_\mY^\top \mY \mV_{\greek{\Omega}} \Sigma^* \mU_{\greek{\Omega},1}^\top \widehat{\mU}_\mY, \quad   \mB_{\text{R}} = \widehat{\mU}_\mY^\top \mY \mV_{\greek{\Omega}} \Sigma^* \mU_{\greek{\Omega},2}^\top,
    \label{eqn:Ar_Br_limit}
\end{equation}
where $[\mU_{\greek{\Omega},1}^{\top} \ \  \mU_{\greek{\Omega},2}^{\top}] = \mU_{\greek{\Omega}}^{\top}$ \  with \  $\mU_{\greek{\Omega},1} \in \R^{d_\vx \times (d_\vx+d_\vu)}, \mU_{\greek{\Omega},2} \in \R^{d_\vu \times (d_\vx+d_\vu)}$, and \\
$\Sigma^* = \lim_{\varepsilon \rightarrow 0} (\greek{\Sigma}_{\greek{\Omega}}^\top \mU_{\greek{\Omega},1}^\top \widehat{\mU}_\mY \widehat{\mU}_\mY^\top \mU_{\greek{\Omega},1} \greek{\Sigma}_{\greek{\Omega}} 
    + \greek{\Sigma}_{\greek{\Omega}}^\top \mU_{\greek{\Omega},2}^\top \mU_{\greek{\Omega},2} \greek{\Sigma}_{\greek{\Omega}} + \varepsilon^2 \mI_n)^{-1} \greek{\Sigma}_{\greek{\Omega}}^\top$.

\label{theorem:dmdc_link}
\end{corollary}

\textit{Proof.} See appendix \ref{apdx:proof_dmdc_link}.

\begin{remark}
It can be verified easily that if we use the truncated SVD (as defined by \ref{eqn:svd}), instead of the full SVD, for $\greek{\Omega}$ in corollary \ref{theorem:dmdc_link}, we get an approximation of (\ref{eqn:Ar_Br_limit}):
\begin{equation}
    \widehat{\mA}_{\text{R}} = \widehat{\mU}_\mY^\top \mY \widehat{\mV}_{\greek{\Omega}} \widehat{\Sigma}^* \widehat{\mU}_{\greek{\Omega},1}^\top \widehat{\mU}_\mY, \quad  \widehat{\mB}_{\text{R}} = \widehat{\mU}_\mY^\top \mY \widehat{\mV}_{\greek{\Omega}} \widehat{\Sigma}^* \widehat{\mU}_{\greek{\Omega},2}^\top,
    \label{eqn:Ar_Br_approx}
\end{equation}
where $\widehat{\Sigma}^* = \lim_{\varepsilon \rightarrow 0} (\widehat{\greek{\Sigma}}_{\greek{\Omega}}^\top \widehat{\mU}_{\greek{\Omega},1}^\top \widehat{\mU}_\mY \widehat{\mU}_\mY^\top \widehat{\mU}_{\greek{\Omega},1} \widehat{\greek{\Sigma}}_{\greek{\Omega}} 
    + \widehat{\greek{\Sigma}}_{\greek{\Omega}}^\top \widehat{\mU}_{\greek{\Omega},2}^\top \widehat{\mU}_{\greek{\Omega},2} \widehat{\greek{\Sigma}}_{\greek{\Omega}} + \varepsilon^2 \mI_{r_{\vx\vu}})^{-1} \widehat{\greek{\Sigma}}_{\greek{\Omega}}^\top$. We can see that (\ref{eqn:Ar_Br_approx}) has the same form as (\ref{eqn:dmdc-params}b), except $\greek{\widehat{\Sigma}}_{\greek{\Omega}}^{-1}$ is replaced with $\widehat{\Sigma}^*$.
\end{remark}

All the aforementioned results are derived for a fixed $\mE_\vx$ and the relation to the DMDc is specific to the case $\mE_\vx = \widehat{\mU}_\mY^\top$. 
Note that the columns of the $\widehat{\mU}_\mY$ are the left singular vectors, corresponding to the leading singular values, of $\mY$. Equivalently, those are also the eigenvectors, corresponding to the leading eigenvalues, of the covariance matrix $\mY \mY^\top$. $L_{\text{pred}}$ alone does not constrain $\mE_\vx$ to take a similar form and we need another loss term to encourage such form for the encoder. To this end, we build on the findings in  \cite{baldi1989neural}, which explore the similarity between principle component analysis and linear autoencoders optimized with the objective function:
    $L_{\text{recon}}  (\mE_\vx, \mD_\vx) = \frac{1}{n} \sum_{i=1}^{n} \big\|\vx(t_i) -  \mD_\vx \mE_\vx \vx(t_i) \big\|^2$.
It is shown in \cite{baldi1989neural} that all the critical points of $L_{\text{recon}} $ correspond to projections onto subspaces associated with subsets of eigenvectors of the covariance matrix $\mY \mY^\top$. Moreover, $L_{\text{recon}}$ has a unique global minimum corresponding to the first $r_\vx$ (i.e., the desired dimension of the reduced state) number of eigenvectors of $\mY \mY^\top$, associated with the leading $r_\vx$ eigenvalues. In other words, for any invertible matrix $\mC \in \R^{r_\vx \times r_\vx}$, $\mD_\vx = \mU_{r_\vx} \mC$ and
$\mE_\vx = \mC^{-1} \mU_{r_\vx}^\top$ globally minimizes $L_{\text{recon}} $, where $\mU_{r_\vx}$ denotes the matrix containing leading $r_\vx$ eigenvectors of
$\mY \mY^\top$. Since the left singular vectors of $\mY$ are the eigenvectors of $\mY \mY^\top$, we have $\mU_{r_\vx} = \widehat{\mU}_\mY$.
Hence, we consider to utilize $L_{\text{recon}}$ to promote learning an encoder $\mE_\vx$  in the form of $\mC^{-1} \widehat{\mU}_\mY^\top$. 
Accordingly, we propose to minimize the following objective function to encourage a DMDc-like solution for LAROM:
\begin{equation}
    L  (\mE_\vx, \mD_\vx, \mG) = L_{\text{pred}}  (\mE_\vx, \mG) + \beta_1 L_{\text{recon}}  (\mE_\vx, \mD_\vx),
    \label{eqn:lin_losses}
\end{equation}
where $\beta_1 > 0$ is a tunable hyperparameter. 

It is important to note that $L_{\text{recon}}$ is minimized for any invertible matrix $\mC$, $\mD_\vx = \widehat{\mU}_\mY \mC$, and $\mE_\vx = \mC^{-1} \widehat{\mU}_\mY^\top$. When optimized using gradient descent, it is highly unlikely to get $\mC$ as the identity matrix like DMDc. Rather, we expect a random $\mC$. Therefore, we need additional constraints to promote similarity with DMDc. For this purpose, we tie the matrices $\mE_\vx$ and $\mD_\vx$ to be the transpose of each other and add a semi-orthogonality constraint $\beta_4 \| \mE_\vx \mE_\vx^\top - \mI_{r_x}\|, \beta_4 > 0$ to the optimization objective of (\ref{eqn:lin_losses}).

\begin{figure}[t]
\begin{center}
\begin{subfigure}[b]{0.5\textwidth}
    \includegraphics[width=0.7\linewidth]{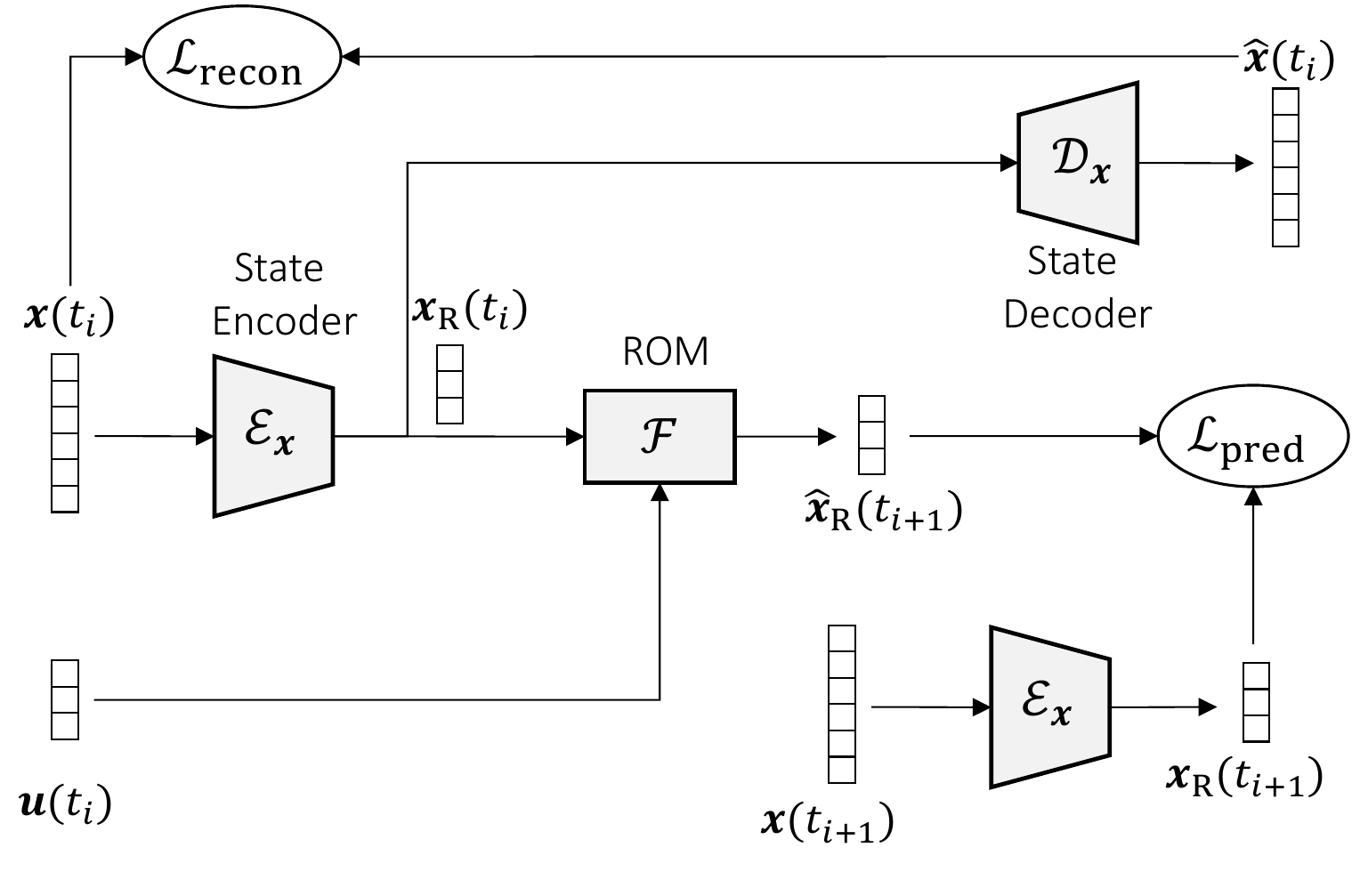}
    \caption{}
    \label{fig:autoencoder}
\end{subfigure}
\begin{subfigure}[b]{0.5\textwidth}
    \includegraphics[width=0.8\linewidth]{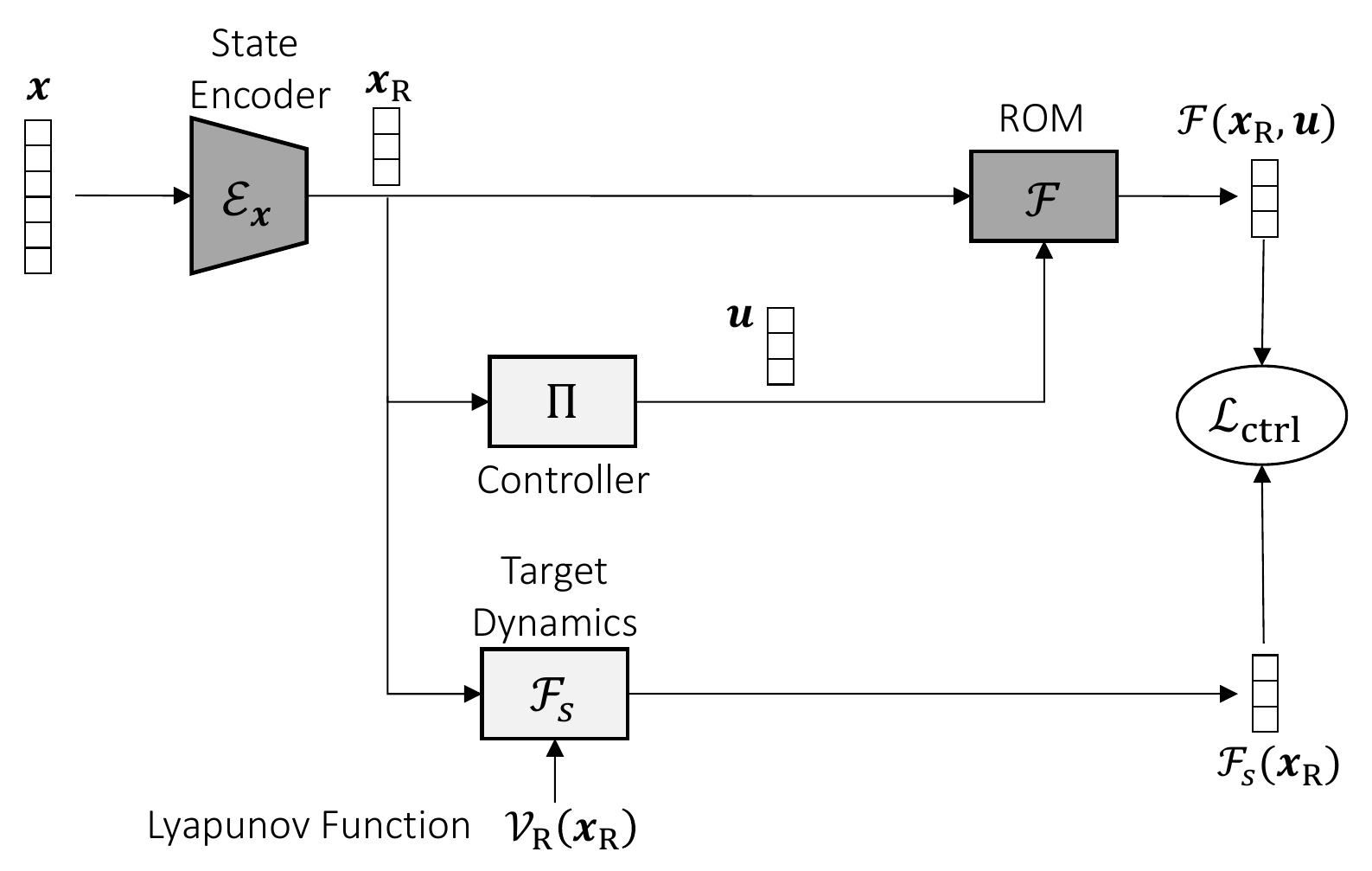}
    \caption{}
    \label{fig:nroc}
\end{subfigure}
\end{center}
\vspace{-10pt}
\caption{(a): Autoencoding architecture for reduced-order modeling. The state encoder $\mathcal{E}_\vx$ reduces the dimension of the state. The ROM $\mathcal{F}$ takes the current reduced state and actuation to predict the next reduced state, which is then uplifted to the full state by the state decoder $\mathcal{D}_\vx$. All modules are trained together using a combined loss involving  $\mathcal{L}_{\text{pred}}$ and $\mathcal{L}_{\text{recon}}$. (b): The control learning process. Given a reduced state, $\mathcal{F}_s$ predicts a target dynamics for the closed-loop system, and the controller $\Pi$ predicts an actuation to achieve that target. Both the modules are trained jointly using the loss function $\mathcal{L}_{\text{ctrl}}$. Parameters of the dark-shaded modules are kept fixed during this process.}
\vspace{-10pt}
\end{figure}

\subsubsection{Extending the linear model to a deep model}
\label{subsub:extend}
Here, we discuss the process of extending LAROM to a nonlinear reduced-order modeling framework.
We replace all the trainable components of LAROM, i.e., $\mE_\vx, \mD_\vx$, and $\mG$, with DNNs. Specifically, we use an encoding function or \textit{encoder} $\mathcal{E}_\vx : \sX \rightarrow \R^{r_\vx}$ and a decoding function or \textit{decoder} $\mathcal{D}_\vx: \R^{r_\vx} \rightarrow \sX$ to transform the high-dimensional system state to low-dimensional features and reconstruct it back, respectively, i.e.,
    $\vx_\text{R} = \mathcal{E}_\vx(\vx), \quad \hat{\vx} = \mathcal{D}_\vx(\vx_\text{R})$,
where $\vx_\text{R} \in \R^{r_\vx}$ denotes the reduced state, and $\hat{\vx}$ is the reconstruction of $\vx$. 
The encoded state and control are then fed to another DNN that represents the reduced order dynamics
    $\frac{d\vx_\text{R}}{dt} = \mathcal{F}(\vx_\text{R}, \vu)$,
where $\mathcal{F}: \R^{r_\vx} \times \R^{d_\vu} \rightarrow \R^{r_\vx}$. Given the current reduced state $\vx_\text{R}(t_i)$ and control input $\vu(t_i)$, the next reduced state $\vx_\text{R}(t_{i+1})$ can be computed by integrating $\mathcal{F}$ using a numerical integrator: 
    $\vx_\text{R}(t_{i+1}) = \vx_\text{R}(t_i) +  \int_{t_i}^{t_{i+1}} \mathcal{F}\big(\vx_\text{R}(t), \vu(t)\big) dt \eqdef \mathcal{G}\big(\vx_\text{R}(t_i), \vu(t_i)\big)$.
We can say that $\mathcal{G}$ is the nonlinear counterpart of $\mG$.

Note, here the ROM is represented as a continuous-time dynamics, unlike the linear case where we used a discrete-time model. 
We use a discrete-time formulation for LAROM to establish its similarity with DMDc, which is formulated in discrete time. DeepROM can be formulated in a similar fashion as well. However,
the specific control learning algorithm we used, which will be discussed in the next subsection, requires vector fields of the learned ROM for training. Therefore, we formulate the ROM in continuous time so that it provides the vector field  $\mathcal{F}(\vx_\text{R}, \vu)$ of the dynamics. In cases where only the prediction model is of interest and control learning is not required, a discrete-time formulation should be used for faster training of the ROM.

We train $\mathcal{E}_\vx, 
\mathcal{D}_\vx$, and $\mathcal{F}$ by minimizing the following loss function, analogous to (\ref{eqn:lin_losses}),
\begin{equation}
    \mathcal{L}(\mathcal{E}_\vx, \mathcal{D}_\vx, \mathcal{F}) = \mathcal{L}_{\text{pred}}(\mathcal{E}_\vx,  \mathcal{F}) + \beta_2 \mathcal{L}_{\text{recon}}(\mathcal{E}_\vx, \mathcal{D}_\vx),
    \label{eqn:loss}
\end{equation}
where $\beta_2 > 0$ is a tunable hyperparameter. $\mathcal{L}_{\text{pred}}$ and $\mathcal{L}_{\text{recon}}$ are defined as 
    $\mathcal{L}_{\text{pred}}(\mathcal{E}_\vx,  \mathcal{F}) = \frac{1}{n} \sum_{i=0}^{n-1} \Big\|\mathcal{E}_\vx\big(\vx(t_{i+1})\big) - \mathcal{G} \Big(\mathcal{E}_\vx\big(\vx(t_i)\big), \vu(t_i)\Big) \Big\|^2$ 
    and
    $\mathcal{L}_{\text{recon}}(\mathcal{E}_\vx, \mathcal{D}_\vx) = \frac{1}{n} \sum_{i=1}^{n} \big\|\vx(t_i) -  \mathcal{D}_\vx \circ  \mathcal{E}_\vx\big(\vx(t_i)\big)\big\|^2$.
Here, the operator $\circ$ denotes the composition of two functions. In experiments, $\mathcal{L}_{\text{recon}}$ also includes the reconstruction loss of the desired state where we want to stabilize the system. Figure \ref{fig:autoencoder} shows the overall framework for training DeepROM.

\subsection{Learning control}
\label{subsec:roc}
Once we get a trained ROM of the form $\frac{d\vx_\text{R}}{dt} = \mathcal{F}(\vx_\text{R}, \vu)$ using the method proposed in section \ref{subsec:rom}, the next goal is to design a controller for the system utilizing that ROM. Since our ROM is represented by DNNs, we need a data-driven method to develop the controller. We adopt the approach presented in \cite{saha2021neural} for learning control laws for nonlinear systems, represented by DNNs. The core idea of the method is to hypothesize a target dynamics that ensures exponentially stability 
of the desired state, and simultaneously learn a control policy to realize that target dynamics in the closed loop. A DNN is used to represent the vector field $\mathcal{F}_s : \R^{r_\vx} \rightarrow \R^{r_\vx}$ of the target dynamics $\frac{d \vx_\text{R}}{dt} = \mathcal{F}_s (\vx_\text{R})$. We use another DNN to represent a controller $\Pi : \R^{r_\vx} \rightarrow \R^{d_\vu}$ that provides the necessary actuation for a given reduced state $\vx_\text{R}$:
    $\vu = \Pi(\vx_\text{R})$.
Finally, the reduced state $\vx_\text{R}$ and actuation $\vu$ are fed to the (trained) ROM of $\frac{d\vx_\text{R}}{dt} = \mathcal{F}(\vx_\text{R}, \vu)$ to get $\mathcal{F}(\vx_\text{R}, \vu)$. The overall framework for learning control is referred to as \textit{deep reduced-order control} (DeepROC) and is shown in Figure \ref{fig:nroc}.

Our training objective is to minimize the difference between $\mathcal{F}(\vx_\text{R}, \vu)$ and $\mathcal{F}_s (\vx_\text{R})$, i.e.,
\begin{align}
    \mathcal{L}_{\text{ctrl}}(\mathcal{F}_s, \Pi)  = \frac{1}{n} \sum_{i=1}^{n} \big\|\mathcal{F} \big(\mathcal{E}_\vx(\vx(t_i)), 
    \Pi \circ \mathcal{E}_\vx(\vx(t_i))\big) \nonumber \\ -  \mathcal{F}_s \circ  \mathcal{E}_\vx\big(\vx(t_i)\big)\big\|^2.
    \label{eqn:control_loss}
\end{align}
To minimize the control effort, we add a regularization loss with (\ref{eqn:control_loss}), and the overall training objective for learning control is given by 
\begin{equation}
    \mathcal{L}_{\text{ctrl,reg}}(\mathcal{F}_s, \Pi) = \mathcal{L}_{\text{ctrl}}(\mathcal{F}_s, \Pi) + \beta_3 \ \frac{1}{n} \sum_{i=1}^{n} \big\| \Pi (\vx_\text{R}(t_i)) \big\|^2,
    \label{eqn:control_reg_loss}
\end{equation}
where $\beta_3 > 0$ is a tunable hyperparameter. 
Here we jointly train the DNNs representing $\Pi$ and $\mathcal{F}_s$ only, whereas the previously-trained DNNs for $\mathcal{E}_\vx$ and $\mathcal{F}$ are kept frozen. Once all the DNNs are trained, we only need $\mathcal{E}_\vx$ and $\Pi$ during evaluation to generate actuation for the actual system, given a full-state observation:
   $\vu = \Pi \circ \mathcal{E}_\vx (\vx) = \pi(\vx)$.
As we mentioned earlier, we require that the target dynamics, hypothesized by a DNN, 
ensures exponential stability of the desired state.
Without loss of generality, we consider stability at $\vx_\text{R} = \mathbf{0}$. The system can be stabilized at any desired state by adding a feedforward component to the control. 
Dynamics represented by a standard neural network does not generally guarantee stability of any equilibrium point.
However, it is possible to design a DNN, by means of Lyapunov functions, to represent a dynamics that ensures exponentially stability of an equilibrium point, as demonstrated in \cite{kolter2019learning}. Accordingly, we represent our target dynamics as follows:
\begin{align}
    &\frac{d \vx_\text{R}}{dt} = \mathcal{F}_s (\vx_\text{R}) = \nonumber \\ & \mathcal{P}(\vx_\text{R}) - \frac{\text{ReLU}\big(\nabla \mathcal{V}_\text{R}(\vx_\text{R})^\top \mathcal{P}(\vx_\text{R}) + \alpha \mathcal{V}_\text{R}(\vx_\text{R})  \big)}{\|\nabla \mathcal{V}_\text{R}(\vx_\text{R}) \|^2} \nabla \mathcal{V}_\text{R}(\vx_\text{R}),
   \label{eqn:exp_stable}
\end{align}
where $\alpha$ is a positive constant, $\text{ReLU}(z) = \max(0, z), \ z \in \R$, $\mathcal{P} : \R^{r_\vx} \rightarrow \R^{r_\vx}$ is an arbitrary function represented by a DNN, and $\mathcal{V}_\text{R}: \R^{r_\vx} \rightarrow \R$ is a candidate Lyapunov function. 
We use 
\begin{equation}
    \mathcal{V}_\text{R}(\normalfont \vx_\text{R}) = \vx_\text{R}^\top \mK \vx_\text{R},
    \label{eqn:lyapunov_function}
\end{equation}
where $\mK \in \R^{r_\vx \times r_\vx}$ is a positive definite matrix.
It is shown in \cite{kolter2019learning} that the origin is exponentially stable for the target dynamics of (\ref{eqn:exp_stable}) for any arbitrary DNN $\mathcal{P}$.
\section{Numerical Experiments}
\label{sec:results}

We consider the Newell-Whitehead-Segel reaction-diffusion equation
\begin{align}
    & \frac{\partial q}{\partial t} = \sigma \nabla^2 q + q (1 - q^2) + \1_{\sW} w \quad \text{in} \  \sI \times \R^+, \nonumber \\
    & \nabla q (\zeta_l, t) =  \nabla q (\zeta_r, t) = 0, \quad t \in \R^+,
    \label{eqn:NWS}
\end{align}
which is used to describe various nonlinear physical systems including Rayleigh-Bénard convection.
In (\ref{eqn:NWS}), $q (\zeta, t) \in \R$ denotes the measurement variable such as concentration or temperature at location $\zeta \in \sI \subset \R$ and time $t$; $\sigma$ denotes the diffusion coefficient; $w(t) \in \R$ is the actuation at time $t$ and $\1_\sW (\zeta)$ is the indicator function with $\sW \subset \sI$; $\zeta_l$ and $\zeta_r$ denote the boundary points of $\sI$.
The considered system is a bistable system with $\pm1$ as stable and $0$ as unstable equilibria. For the control task, we consider feedback stabilization of this system at the unstable equilibrium $0$, as studied in \cite{kalise2018polynomial}. We use $\sI = (-1, 1), \sW = (-0.2, 0.2)$, and $\sigma = 0.2$. Details on dataset generation, neural network architectures, and training settings are given in appendix \ref{apdx:rec-dif}.    

The prediction performance of DeepROM is compared against DMDc and the Deep Koopman model \cite{morton2018deep}. 
The Deep Koopman model shares a similar DNN-based autoencoding structure as ours, with the distinction that its (reduced-order) dynamic model is linear.
The method proposed in \cite{morton2018deep} considers a model predictive scenario, where the state/system matrix of the linear reduced-order model is updated with online observations during operation while the input/control matrix is kept fixed. However, in contrast to the original method, we keep both matrices fixed during the control operation as we consider offline control design in this paper. For the same reason, we  apply linear quadratic regulator (LQR) on the ROM obtained from the Deep Koopman method, instead of model predictive control, to compare the control performance with our method: DeepROC. The control performance is also compared against the reduced-order controller obtained by applying LQR on the ROM derived from DMDc.

\begin{figure}[t]
\begin{center}
\includegraphics[width=1\linewidth]{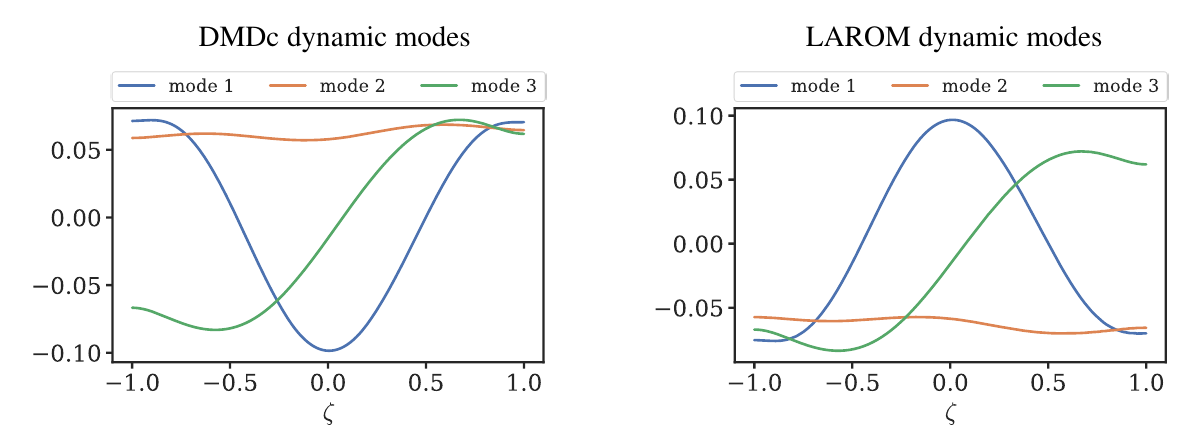}
\end{center}
\vspace{-10pt}
\caption{The first three dynamic modes of the reaction–diffusion system, obtained using DMDc and LAROM.}
\label{fig:NWS_modes}
\vspace{-10pt}
\end{figure}

\begin{figure*}[t]
\begin{center}
\includegraphics[width=0.7\linewidth]{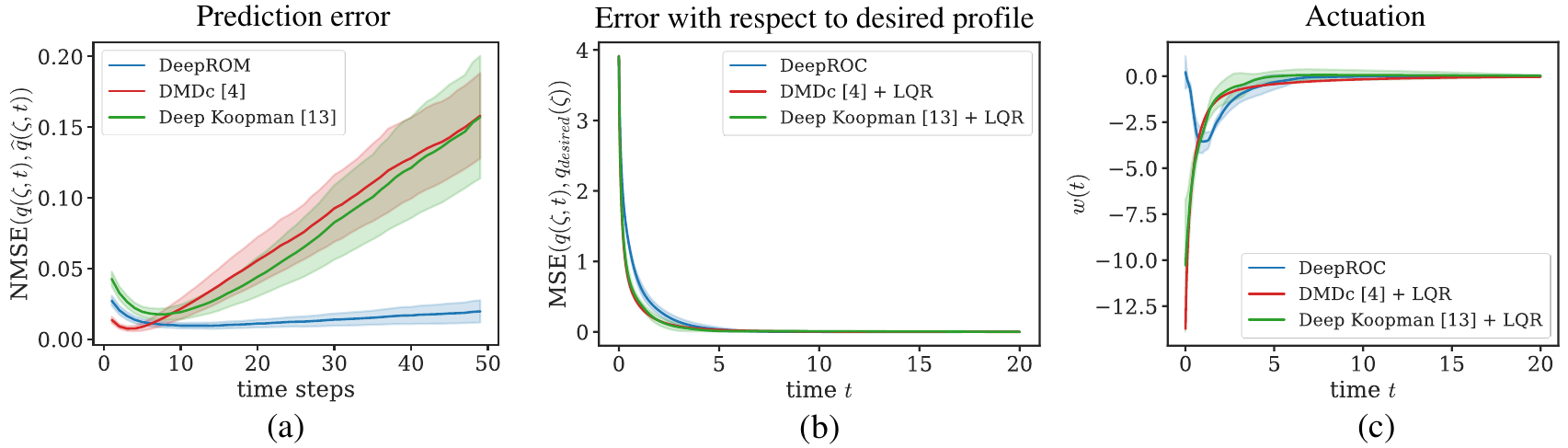}
\end{center}
\vspace{-10pt}
\caption{(a): Prediction performance of 
different methods in the reaction–diffusion example. 
The shaded interval shows the $95\%$ confidence interval around the mean from $100$ test sequences and $3$ different training instances.
(b,c): Control performance of 
different methods in the reaction–diffusion example. %
The shaded interval shows the $1$-standard deviation range around the mean from $3$ different training instances.}
\label{fig:NWS_metrics}
\vspace{-10pt}
\end{figure*}

\begin{figure}[t]
\begin{center}
\includegraphics[width=0.9\linewidth]{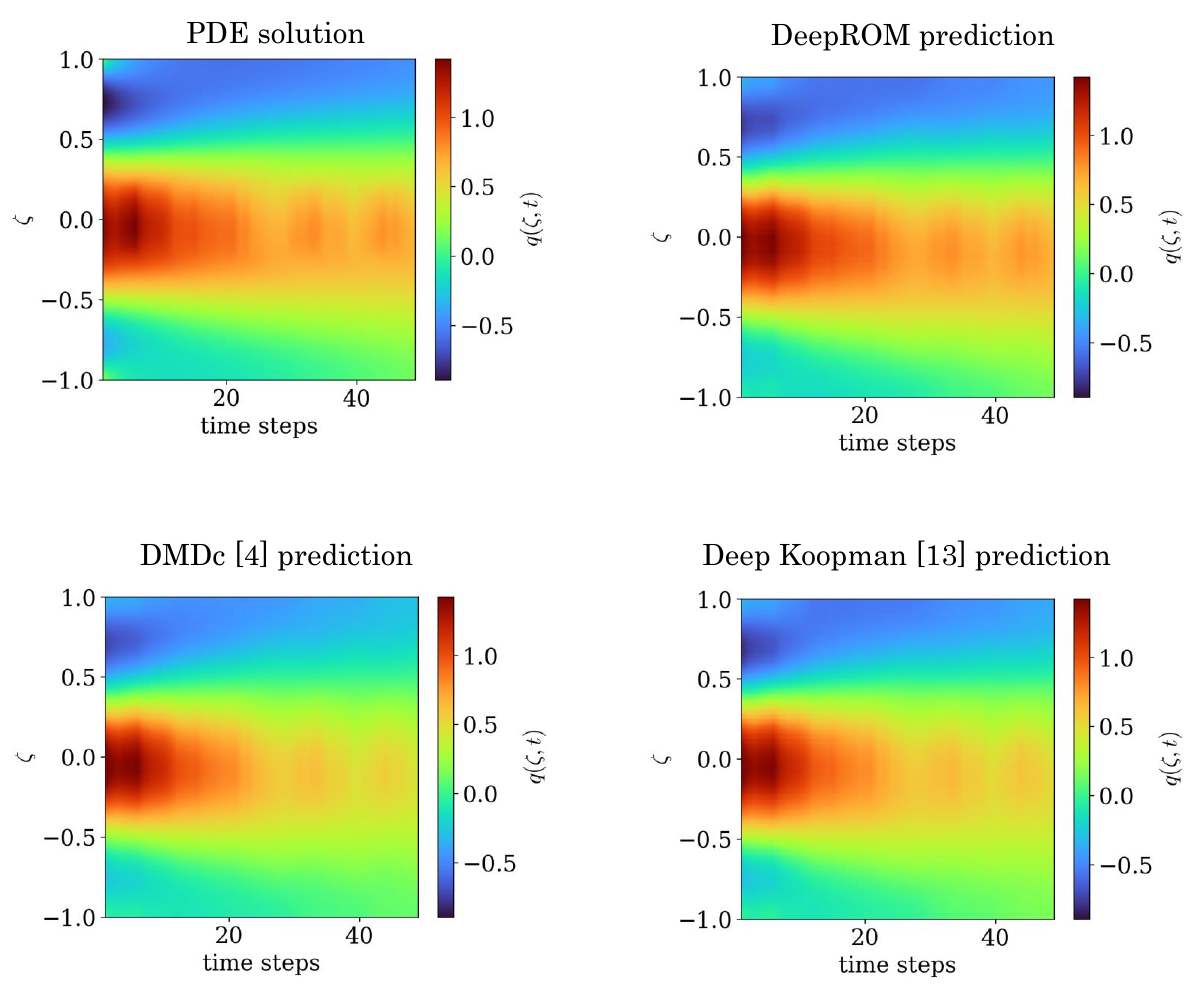}
\end{center}
\vspace{-10pt}
\caption{Qualitative comparison of prediction performance of DMDc, Deep Koopman, and DeepROM for the reaction–diffusion system using one example sequence.}
\label{fig:NWS_pred}
\vspace{-10pt}
\end{figure}

\begin{figure}[t]
\begin{center}
\includegraphics[width=0.9\linewidth]{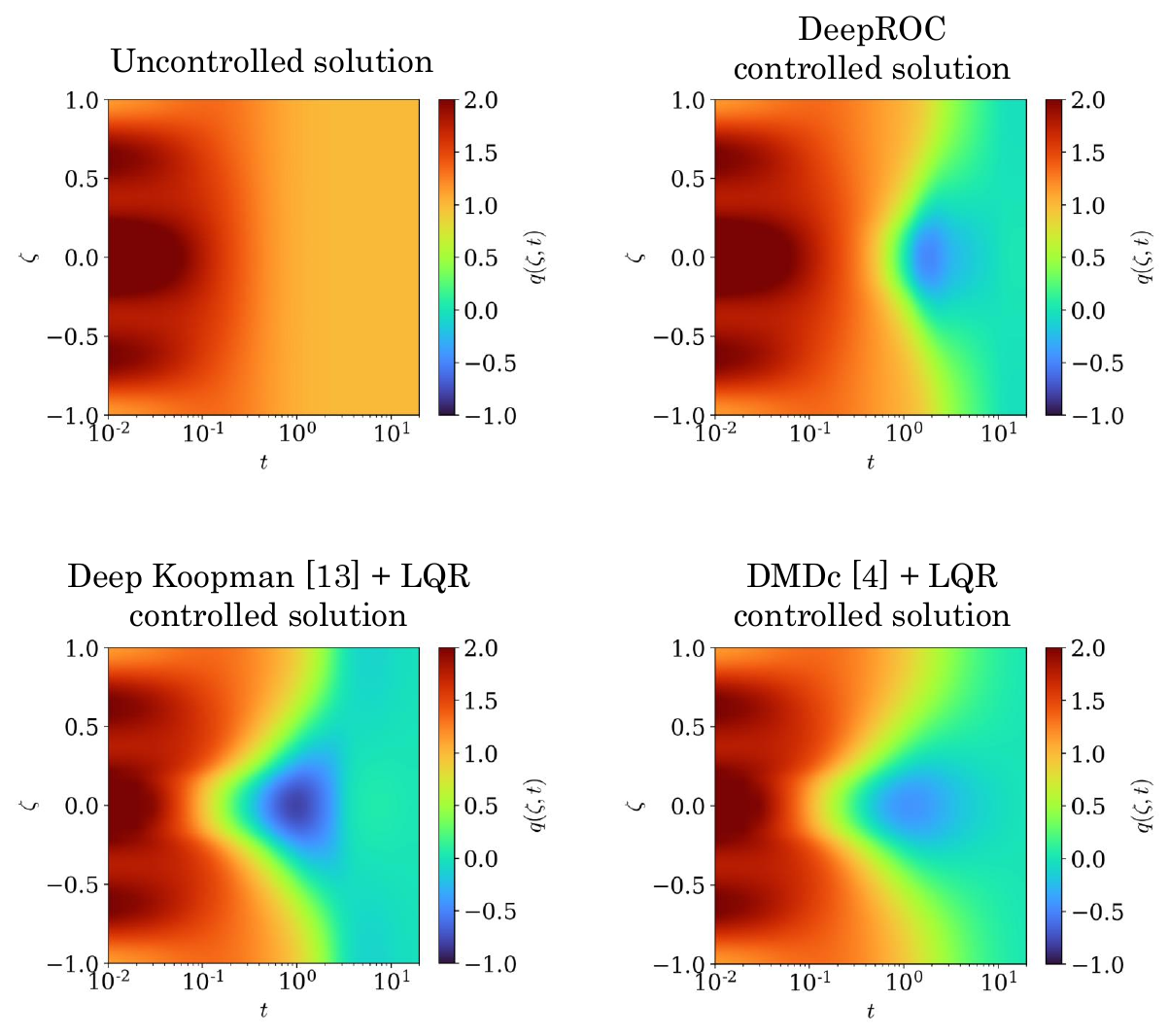}
\end{center}
\vspace{-10pt}
\caption{Visual comparison of the uncontrolled solution and the controlled solutions of the reaction–diffusion system using DeepROC, Deep Koopman + LQR, and DMDc + LQR.}
\label{fig:NWS_visual}
\vspace{-10pt}
\end{figure}

\textbf{Similarity with DMDc.}
To investigate the similarity with DMDc, we first train LAROM using gradient descent to minimize the objective (\ref{eqn:lin_losses}) with the semi-orthogonality regularization and enforcing $\mD_\vx = \mE_\vx^\top$, as discussed in \ref{subsub:connection}. The dynamic modes for LAROM are computed as $\greek{\varphi}_i = \mD_\vx \vz_i$, where $\vz_i$ is the $i^{\text{th}}$ eigenvector of $\mA_{\normalfont\text{R}}$. Similarly, the dynamic modes for DMDc are computed as $\greek{\varphi}_{i, \normalfont\text{DMDc}} = \mD_{\normalfont\text{DMDc}} \vz_{i,\normalfont\text{DMDc}}$, where $\vz_{i,\normalfont\text{DMDc}}$ is the $i^{\text{th}}$ eigenvector of $\mA_{\normalfont\text{R,DMDc}}$. Note, these dynamic modes are similar to the ones used in the original DMD algorithm \cite{schmid2010dynamic}, not the exact modes obtained in \cite{proctor2016dynamic}. Exact modes cannot be computed for LAROM since it does not involve SVD. Modes defined by $\greek{\varphi}_{i, \normalfont\text{DMDc}} = \mD_{\normalfont\text{DMDc}} \vz_{i,\normalfont\text{DMDc}} = \widehat{\mU}_\mY \vz_{i,\normalfont\text{DMDc}}$ are the orthogonal projection of the exact modes onto the range of $\mY$ (Theorem 3, \cite{tu2013dynamic}).
Fig. \ref{fig:NWS_modes} compares the dynamic modes obtained using DMDc and LAROM for the case when the dimension of the ROMs is $3$. It is important to note that the numbering of the modes is arbitrary as the optimal ranking of DMDc modes is not trivial. The correspondence between the DMDc modes and LAROM modes are determined by comparing the eigenvalues of $\mA_{\normalfont\text{R,DMDc}}$ and $\mA_{\normalfont\text{R}}$. Dynamic modes of both methods are similar except for the different signs of the first two modes.   

\textbf{Prediction performance.}
We compare the performance of DeepROM, 
Deep Koopman model, and DMDc in the prediction task. 
Fig. \ref{fig:NWS_metrics}(a) shows the quantitative 
comparison of the recursive multi-step predictions obtained using DMDc, Deep Koopman model, and DeepROM. The prediction error is computed as \textit{normalized mean squared error} (NMSE) with respect to the solution obtained using the PDE solver. 
Prediction error increases more quickly for DMDc and Deep Koopman than DeepROM
as the linear ROMs become less accurate in the long term. 
A qualitative comparison of the prediction performance of the methods for an example sequence is shown in Fig. \ref{fig:NWS_pred}.

\textbf{Control performance.} 
Fig. \ref{fig:NWS_metrics}(b,c) show the control performance of DeepROC, Deep Koopman + LQR, and  
DMDc + LQR in the task of stabilizing the system at the unstable equilibrium 0 from an initial state $2 + \cos(2 \pi \zeta) \cos(\pi \zeta)$. We use the following two metrics for comparison: 
\begin{enumerate}[(i)]
    \item mean squared error over time between the controlled solutions and the desired profile
    \item the amount of actuation applied
\end{enumerate}
All methods show similar closed-loop error profiles. However, DeepROC requires significantly less amount of actuation in comparison with DMDc + LQR and Deep Koopman + LQR to reach a similar steady-state error. DeepROC can account for the decaying nonlinear term $- q^3$ present in the system (\ref{eqn:NWS}) and therefore learns to apply less actuation. 
A qualitative comparison of the uncontrolled solution and the controlled solutions obtained using the three methods is shown in Fig. \ref{fig:NWS_visual}. 

\section{Conclusion}
\label{sec:conclusion}

We presented a framework for autoencoder-based modeling and control learning for PDE-driven dynamical systems. 
The proposed reduced-order modeling framework is grounded on the connection between dynamic mode decomposition for controlled systems and a linear autoencoding architecture that can be trained using gradient descent.   
As demonstrated in our numerical experiments, DeepROM offers better prediction accuracy than a linear ROM over a relatively longer prediction horizon when applied to a nonlinear system. Additionally, DeepROC requires less amount of actuation to reach the desired state for such a system.

\vspace{5mm}
\appendices
\section{Proofs}

\subsection{Proof of theorem \ref{theorem:lin_opt}}
\label{apdx:proof_lin_opt}

We can write $L_{\text{pred}} (\mE_\vx, \mG)$ as follows,
\begin{align}
     L_{\text{pred}} (\mE_\vx, \mG) &= \frac{1}{n} \sum_{i=0}^{n-1} \big\|\mE_\vx \vx(t_{i+1}) - \mG \mE_{\vx \vu} \bm{\omega}(t_i) \big\|^2 \nonumber \\
     &= \frac{1}{n} \big\| \vecm(\mE_\vx \mY) - \vecm(\mG \mE_{\vx \vu} \greek{\Omega}) \big\|^2 \nonumber \\
     &= \frac{1}{n}\big\| \vecm(\mE_\vx \mY) - (\greek{\Omega}^\top \mE_{\vx \vu}^\top \otimes \mI_{r_\vx}) \vecm(\mG) \big\|^2,
     \label{eqn:lin_loss_pred_vec}
\end{align}
where $\otimes$ and $\vecm(\cdot)$ denotes the Kronecker product and vectorization of matrices, respectively. 
The third equality is obtained using properties of the Kronecker product. Applying the principles of linear least squares optimization, we obtain the following outcome. For fixed $\mE_\vx$, (\ref{eqn:lin_loss_pred_vec}) is convex in coefficient of $\mG$, and
$\mG$ corresponds to a global minimum of $L_{\text{pred}}$ if and only if 
\begin{align}
    (\greek{\Omega}^\top \mE_{\vx \vu}^\top \otimes \mI_{r_\vx})^\top (\greek{\Omega}^\top \mE_{\vx \vu}^\top \otimes \mI_{r_\vx}) \vecm(\mG) \nonumber \\ = (\greek{\Omega}^\top \mE_{\vx \vu}^\top \otimes \mI_{r_\vx})^\top \vecm(\mE_\vx \mY).
    \label{eqn:lin_loss_pred_vec_sol}
\end{align}
Using the properties of the Kronecker product, we can write (\ref{eqn:lin_loss_pred_vec_sol}) as
\begin{equation}
    (\mE_{\vx \vu} \greek{\Omega} \greek{\Omega}^\top \mE_{\vx \vu}^\top  \otimes \mI_{r_\vx}) \vecm(\mG) = (\mE_{\vx \vu} \greek{\Omega} \otimes \mI_{r_\vx}) \vecm(\mE_\vx \mY),
    \label{eqn:lin_loss_pred_vec_sol2}
\end{equation}
which yields $\mG \mE_{\vx \vu} \greek{\Omega} \greek{\Omega}^\top \mE_{\vx \vu}^\top = \mE_\vx \mY \greek{\Omega}^\top \mE_{\vx \vu}^\top$, i.e., (\ref{eqn:lin_loss_pred_sol_1}). 

If $\mE_{\vx}$ has full rank $r_\vx$, then $\mE_{\vx \vu} = \begin{bmatrix}
\mE_\vx & \mathbf{0}\\
\mathbf{0} & \mI_{d_\vu}
\end{bmatrix} \in \R^{(r_\vx + d_\vu) \times (d_\vx + d_\vu)}$ has full rank $(r_\vx + d_\vu)$. If $\greek{\Omega} \greek{\Omega}^\top \in \R^{(d_\vx + d_\vu) \times (d_\vx + d_\vu)}$ is non-singular, then $\greek{\Omega}$ has full row-rank $(d_\vx + d_\vu)$. Consequently, applying the properties of matrix rank, we have
\begin{align}
    \rank(\mE_{\vx \vu} \greek{\Omega} \greek{\Omega}^\top \mE_{\vx \vu}^\top) &=  \rank(\mE_{\vx \vu} \greek{\Omega}) \nonumber \\ &= \rank(\mE_{\vx \vu}) = r_\vx + d_\vu.
\end{align}
Hence, the symmetric positive semidefinite matrix $\mE_{\vx \vu} \greek{\Omega} \greek{\Omega}^\top \mE_{\vx \vu}^\top$ has full rank and is therefore positive definite. Using the properties of the Kronecker product of positive definite matrices, we  
can see that $(\greek{\Omega}^\top \mE_{\vx \vu}^\top \otimes \mI_{r_\vx})^\top (\greek{\Omega}^\top \mE_{\vx \vu}^\top \otimes \mI_{r_\vx}) = (\mE_{\vx \vu} \greek{\Omega} \greek{\Omega}^\top \mE_{\vx \vu}^\top  \otimes \mI_{r_\vx})$ is positive definite as well. Therefore,   
(\ref{eqn:lin_loss_pred_vec}) is strictly convex in the coefficients of $\mG$ and has a unique minimum. 
Since $\mE_{\vx \vu} \greek{\Omega} \greek{\Omega}^\top \mE_{\vx \vu}^\top \succ 0$, it is invertible. Hence, from (\ref{eqn:lin_loss_pred_sol_1}), we can say that the unique minimum of (\ref{eqn:lin_loss_pred_vec}) is reached at $\mG = \mE_\vx \mY \greek{\Omega}^\top \mE_{\vx \vu}^\top (\mE_{\vx \vu} \greek{\Omega} \greek{\Omega}^\top \mE_{\vx \vu}^\top)^{-1}$, i.e., (\ref{eqn:lin_loss_pred_sol_2}). \hspace*{\fill} $\blacksquare$

\subsection{An alternative representation of (\ref{eqn:lin_loss_pred_sol_2})}
\label{apdx:alt_G}
Here we provide an alternative representation of (\ref{eqn:lin_loss_pred_sol_2}) required to prove corollary \ref{theorem:dmdc_link}.

\begin{lemma}
Consider the (full) SVD of the data matrix $\greek{\Omega}$ given by $\greek{\Omega} = \mU_{\greek{\Omega}} \greek{\Sigma}_{\greek{\Omega}} \mV_{\greek{\Omega}}^\top$, where $\mU_{\greek{\Omega}} \in \R^{(d_\vx+d_\vu) \times (d_\vx+d_\vu)}, \greek{\Sigma}_{\greek{\Omega}} \in \R^{(d_\vx+d_\vu) \times n}$, and $\mV_{\greek{\Omega}} \in \R^{n \times n}$. 
(\ref{eqn:lin_loss_pred_sol_2}) can be expressed as 
\begin{align}
    \mG = \lim_{\varepsilon \rightarrow 0} \mE_\vx \mY \mV_{\greek{\Omega}} (\greek{\Sigma}_{\greek{\Omega}}^\top \mU_{\greek{\Omega}}^\top \mE_{\vx \vu}^\top \mE_{\vx \vu} \mU_{\greek{\Omega}} \greek{\Sigma}_{\greek{\Omega}} \nonumber \\ + \varepsilon^2 \mI_n)^{-1} \greek{\Sigma}_{\greek{\Omega}}^\top \mU_{\greek{\Omega}}^\top \mE_{\vx \vu}^\top.
    \label{eqn:G_alternatives}
\end{align}
\label{lemma:G_alternatives}
\end{lemma}
\textit{Proof}. 
Replacing $\greek{\Omega}$ with its SVD in (\ref{eqn:lin_loss_pred_sol_2}) we get,
\begin{align}
    & \mG = \nonumber \\ &\mE_\vx \mY \mV_{\greek{\Omega}} \greek{\Sigma}_{\greek{\Omega}}^\top \mU_{\greek{\Omega}}^\top \mE_{\vx \vu}^\top (\mE_{\vx \vu} \mU_{\greek{\Omega}} \greek{\Sigma}_{\greek{\Omega}} \mV_{\greek{\Omega}}^\top \mV_{\greek{\Omega}} \greek{\Sigma}_{\greek{\Omega}}^\top \mU_{\greek{\Omega}}^\top \mE_{\vx \vu}^\top)^{-1} \nonumber \\
    &= \mE_\vx \mY \mV_{\greek{\Omega}} \greek{\Sigma}_{\greek{\Omega}}^\top \mU_{\greek{\Omega}}^\top \mE_{\vx \vu}^\top (\mE_{\vx \vu} \mU_{\greek{\Omega}} \greek{\Sigma}_{\greek{\Omega}}  \greek{\Sigma}_{\greek{\Omega}}^\top \mU_{\greek{\Omega}}^\top \mE_{\vx \vu}^\top)^{-1} \nonumber \\
    &= \mE_\vx \mY \mV_{\greek{\Omega}} (\mE_{\vx \vu} \mU_{\greek{\Omega}} \greek{\Sigma}_{\greek{\Omega}})^+,
    \label{eqn:G_alternative_1}
\end{align}
where $(\cdot)^+$ denotes the Moore-Penrose inverse of a matrix. 
The second equality is due to the orthogonality of $\mV_{\greek{\Omega}}$. Substituting $(\mE_{\vx \vu} \mU_{\greek{\Omega}} \greek{\Sigma}_{\greek{\Omega}})^+$ with 
the \textit{limit definition} (see \cite{albert1972regression}) of the Moore-Penrose inverse, we get  
\begin{align}
    \mG = \lim_{\varepsilon \rightarrow 0} \mE_\vx \mY \mV_{\greek{\Omega}} (\greek{\Sigma}_{\greek{\Omega}}^\top \mU_{\greek{\Omega}}^\top \mE_{\vx \vu}^\top \mE_{\vx \vu} \mU_{\greek{\Omega}} \greek{\Sigma}_{\greek{\Omega}} \nonumber \\ + \varepsilon^2 \mI_n)^{-1} \greek{\Sigma}_{\greek{\Omega}}^\top \mU_{\greek{\Omega}}^\top \mE_{\vx \vu}^\top.
    \label{eqn:G_alternative_3}
\end{align}
\hspace*{\fill} $\blacksquare$

\subsection{Proof of Corollary \ref{theorem:dmdc_link}}
\label{apdx:proof_dmdc_link}

By the definition of truncated SVD, the columns of $\widehat{\mU}_\mY$ are orthonormal. Therefore, $\widehat{\mU}_\mY^\top$ has full row-rank $r_\vx$.
Hence, by theorem \ref{theorem:lin_opt} and lemma \ref{lemma:G_alternatives}, if $\mE_\vx = \widehat{\mU}_\mY^\top$, and $\greek{\Omega} \greek{\Omega}^\top$ is non-singular, then the unique minimum of $L_{\text{pred}}$,  is reached when 
\begin{align}
    \mG = \widehat{\mU}_\mY^\top \mY \mV_{\greek{\Omega}} (\mE_{\vx \vu} \mU_{\greek{\Omega}} \greek{\Sigma}_{\greek{\Omega}})^+ \nonumber \\
    = \lim_{\varepsilon \rightarrow 0} \widehat{\mU}_\mY^\top \mY \mV_{\greek{\Omega}} (\greek{\Sigma}_{\greek{\Omega}}^\top \mU_{\greek{\Omega}}^\top \mE_{\vx \vu}^\top \mE_{\vx \vu} \mU_{\greek{\Omega}} \greek{\Sigma}_{\greek{\Omega}} \nonumber \\ + \varepsilon^2 \mI_n)^{-1} \greek{\Sigma}_{\greek{\Omega}}^\top \mU_{\greek{\Omega}}^\top \mE_{\vx \vu}^\top.
    \label{eqn:G_with_Uy}
\end{align}

Now, substituting $\mE_\vx = \widehat{\mU}_\mY^\top$ in $\mE_{\vx\vu}$, and using the partition $\mU_{\greek{\Omega}}^{\top} = [\mU_{\greek{\Omega},1}^{\top} \ \  \mU_{\greek{\Omega},2}^{\top}]$, where $\mU_{\greek{\Omega},1} \in \R^{d_\vx \times (d_\vx+d_\vu)}, \mU_{\greek{\Omega},2} \in \R^{d_\vu \times (d_\vx+d_\vu)}$, we get 
\begin{equation}
    \mE_{\vx\vu} \mU_{\greek{\Omega}} = \begin{bmatrix}
    \widehat{\mU}_\mY^\top & \mathbf{0}\\
    \mathbf{0} & \mI_{d_\vu} 
    \end{bmatrix}
    \begin{bmatrix}
    \mU_{\greek{\Omega},1} \\
    \mU_{\greek{\Omega},2} 
    \end{bmatrix}
    = \begin{bmatrix}
    \widehat{\mU}_\mY^\top \mU_{\greek{\Omega},1} \\
    \mU_{\greek{\Omega},2}
    \end{bmatrix},
    \label{eqn:EU}
\end{equation}
and 
\begin{align}
    \mU_{\greek{\Omega}}^\top \mE_{\vx\vu}^\top \mE_{\vx\vu} \mU_{\greek{\Omega}} &= \begin{bmatrix}
    \mU_{\greek{\Omega},1}^\top \widehat{\mU}_\mY &
    \mU_{\greek{\Omega},2}^\top
    \end{bmatrix}
    \begin{bmatrix}
    \widehat{\mU}_\mY^\top \mU_{\greek{\Omega},1} \\
    \mU_{\greek{\Omega},2}
    \end{bmatrix} \nonumber \\
    &= \mU_{\greek{\Omega},1}^\top \widehat{\mU}_\mY \widehat{\mU}_\mY^\top \mU_{\greek{\Omega},1} 
    + \mU_{\greek{\Omega},2}^\top \mU_{\greek{\Omega},2}.
    \label{eqn:UtEtEU}
\end{align}

Plugging (\ref{eqn:EU}) and (\ref{eqn:UtEtEU}) into (\ref{eqn:G_with_Uy}) leads to 
\begin{align}
    \mG = \lim_{\varepsilon \rightarrow 0} \widehat{\mU}_\mY^\top \mY \mV_{\greek{\Omega}} (\greek{\Sigma}_{\greek{\Omega}}^\top \mU_{\greek{\Omega},1}^\top \widehat{\mU}_\mY \widehat{\mU}_\mY^\top \mU_{\greek{\Omega},1} \greek{\Sigma}_{\greek{\Omega}} \nonumber \\
    + \greek{\Sigma}_{\greek{\Omega}}^\top \mU_{\greek{\Omega},2}^\top \mU_{\greek{\Omega},2} \greek{\Sigma}_{\greek{\Omega}} + \varepsilon^2 \mI_n)^{-1} \greek{\Sigma}_{\greek{\Omega}}^\top \begin{bmatrix}
    \mU_{\greek{\Omega},1}^\top \widehat{\mU}_\mY & 
    \mU_{\greek{\Omega},2}^\top
    \end{bmatrix}.
    \label{eqn:G_limit}
\end{align}
Defining $\Sigma^* \eqdef \lim_{\varepsilon \rightarrow 0} (\greek{\Sigma}_{\greek{\Omega}}^\top \mU_{\greek{\Omega},1}^\top \widehat{\mU}_\mY \widehat{\mU}_\mY^\top \mU_{\greek{\Omega},1} \greek{\Sigma}_{\greek{\Omega}} 
    + \greek{\Sigma}_{\greek{\Omega}}^\top \mU_{\greek{\Omega},2}^\top \mU_{\greek{\Omega},2} \greek{\Sigma}_{\greek{\Omega}} + \varepsilon^2 \mI_n)^{-1} \greek{\Sigma}_{\greek{\Omega}}^\top$,
we can split (\ref{eqn:G_limit}) into
\begin{equation*}
    \mA_{\text{R}} = \widehat{\mU}_\mY^\top \mY \mV_{\greek{\Omega}} \Sigma^* \mU_{\greek{\Omega},1}^\top \widehat{\mU}_\mY, \  \text{and} \quad \mB_{\text{R}} = \widehat{\mU}_\mY^\top \mY \mV_{\greek{\Omega}} \Sigma^* \mU_{\greek{\Omega},2}^\top,
\end{equation*}
which is (\ref{eqn:Ar_Br_limit}). \hspace*{\fill} $\blacksquare$

\section{Details on numerical experiment}
\label{apdx:rec-dif}

\subsection{Dataset}
We generate $100$ training sequences of length $50$ (i.e., $n=50$) with time step size $0.01$ and $256$ nodes in $\sI$ using the finite element method. The initial conditions and actuations of these sequences are given by
\begin{equation}
    q(\zeta, 0) = |a| \sum_{k=0}^4 b_k T_k (\zeta), \quad \zeta \in \sI, 
\end{equation}
and
\begin{equation}
    w(t_i) = 10 g_i \max_\zeta |q(\zeta, t_{i-1})|, \quad i = 1, 2, \cdots, 49,
\end{equation}
where $T_k$ denotes the $k^\text{th}$ Chebyshev polynomial of the first kind, and $a \sim \mathcal{N}(0, 1)$, $b_k, g_i \sim \mathcal{U}(-1, 1)$ are chosen randomly. Similarly, $100$ sequences are generated for the test set to evaluate the prediction performance.

\subsection{DNN architectures}
 
The state encoder comprises two 1D convolutional layers, followed by two fully connected (FC) layers. The first convolutional layer has 32 filters of size $3$, utilizing a ReLU activation function and a stride of 2. The second convolutional layer maintains the same configuration but lacks the ReLU activation. The subsequent FC layer contains 64 neurons with ReLU activation. The final or output FC layer is composed of $r_\vx$ neurons and does not use any bias.
The arrangement of layers in the state decoder is inverted compared to that of the encoder, with transposed convolutional layers replacing the convolutional layers.
The ROM is designed by breaking the function $\mathcal{F}$ into two components: $\mathcal{F}\big(\vx_\text{R}, \vu\big) = \mathcal{F}_{\text{auto}}\big(\vx_\text{R}\big) + \mathcal{F}_{\text{forced}}\big(\vx_\text{R}, \vu\big) - \mathcal{F}_{\text{forced}}\big(\vx_\text{R}, \mathbf{0}\big)$. $\mathcal{F}_{\text{auto}}$ represents the autonomous dynamics that does not depend on the actuation, whereas $\mathcal{F}_{\text{forced}}$ is responsible for the impact of actuation on dynamics. We observe this structure results in better performance in experiments than a single neural network representing $\mathcal{F}\big(\vx_\text{R}, \vu\big)$.  
Two multilayer perceptrons (MLPs) are employed to realize $\mathcal{F}_{\text{auto}}$ and $\mathcal{F}_{\text{forced}}$, each consisting of three layers. The first and second layers each comprise 100 neurons with ReLU activation, while the output layer consists of $r_\vx$ neurons.
The output of the ROM is integrated using a numerical integrator to get the next state. 
The target dynamics is represented by an MLP akin to the one employed for $\mathcal{F}_{\text{auto}}$, followed by a stability criterion in the form of (\ref{eqn:exp_stable}). 
The controller is implemented using another MLP with a similar three-layer configuration, while the output layer is adjusted to accommodate the actuation dimension. 

For the encoder and decoder of the Deep Koopman model, we use the identical architectures as those employed in our DeepROM model.

\subsection{Training settings}
We use $r_\vx = 5$ in the prediction task and $r_\vx = 2$ in the control task for all the methods.
All modules are implemented in PyTorch. In both of the learning phases, learning ROM and learning controller, we use the Adam optimizer with an initial learning rate of $0.001$ and apply an exponential scheduler with a decay of $0.99$. Modules are trained for $100$ epochs in mini-batches of size $32$. $10\%$ of the training data is used for validation to choose the best set of models. For DeepROM training, we use $\beta_2 = 1$ in (\ref{eqn:loss}). For learning control, we use $\beta_3=0.2$ in (\ref{eqn:control_reg_loss}), $\alpha=0.2$ in (\ref{eqn:exp_stable}), and $\mK = 0.5\mI_{r_\vx}$ in (\ref{eqn:lyapunov_function}).
Since the learned ROMs from one training instance to another can vary, the hyperparameter pair ($\alpha, \beta_3$) may require re-tuning accordingly.  

For the Deep Koopman model, 
the input matrix is optimized by gradient descent during training along with the encoder-decoder parameters, whereas the system matrix is obtained using linear least-squares regression. The datasets are divided into staggered 32-step sequences for training, and the model is trained by generating recursive predictions over 32 steps following \cite{morton2018deep}. We employ the same learning hyperparameters as our model, adjusting the mini-batch size to 8 to accommodate multistep recursive prediction.

\bibliographystyle{IEEEtran}
\bibliography{references}


\end{document}